\documentclass[useAMS,longauth,usenatbib,usegraphicx,traditabstract]{aa}

\usepackage{graphicx}
\usepackage{natbib}
\usepackage{amsmath}
\usepackage{psfig}
\bibpunct{(}{)}{;}{a}{}{,}

\def\aap{A\&A}

\def\apjl{ApJ}
\def\apjs{ApJS}

\def\herschel{{\it Herschel}}

\def\qir{$q_{\rm IR}$}

\def\lir{$L_{\rm IR}$}
\def\lrad{$L_{\rm radio}$}
\def\l1.4{$L_{\rm 1.4GHz}$}
\def\s1.4{$S_{\rm 1.4GHz}$}
\def\sir{$S_{\rm IR}$}

\def\td{$T_{\rm d}$}
\def\gs{\mathrel{\raise0.35ex\hbox{$\scriptstyle >$}\kern-0.6em
\lower0.40ex\hbox{{$\scriptstyle \sim$}}}}
\def\ls{\mathrel{\raise0.35ex\hbox{$\scriptstyle <$}\kern-0.6em
\lower0.40ex\hbox{{$\scriptstyle \sim$}}}}

\begin{document}

\title{The far-infrared/radio correlation as probed by
\herschel\thanks{\!\!\!\!{\it Herschel} is an ESA space
observatory with science instruments provided by European-led
Principal Investigator consortia and with important participation from
NASA.}}

\titlerunning{What the FIRRC?}

\author{R.\,J.~Ivison\inst{1,2}
\and B.~Magnelli\inst{3}
\and E.~Ibar\inst{1}
\and P.~Andreani\inst{4,5}
\and D.~Elbaz\inst{6}
\and B.~Altieri\inst{7}
\and A.~Amblard\inst{8}
\and V.~Arumugam\inst{2}
\and R.~Auld\inst{9}
\and H.~Aussel\inst{6}
% ...
}

\institute{UK Astronomy Technology Centre, Royal Observatory, Blackford Hill, Edinburgh EH9 3HJ, UK\\
 \email{rji@roe.ac.uk}
\and Institute for Astronomy, University of Edinburgh, Royal Observatory, Blackford Hill, Edinburgh EH9 3HJ, UK
\and Max-Planck-Institut f\"ur Extraterrestrische Physik (MPE), Postfach 1312, 85741, Garching, Germany
\and ESO, Karl-Schwarzschild-Str.\ 2, 85748 Garching bei M\"unchen, Germany
\and INAF -- Osservatorio Astronomico di Trieste, via Tiepolo 11, 34143 Trieste, Italy
\and Laboratoire AIM-Paris-Saclay, CEA/DSM/Irfu -- CNRS -- Universit\'e Paris Diderot, CE-Saclay, pt courrier 131, F-91191 Gif-sur-Yvette, France
\and Herschel Science Centre, European Space Astronomy Centre, Villanueva de la Ca\~nada, 28691 Madrid, Spain
\and Department of Physics \& Astronomy, University of California, Irvine, CA 92697, USA
\and Cardiff School of Physics and Astronomy, Cardiff University, Queens Buildings, The Parade, Cardiff CF24 3AA, UK
% ...
}

\date{Received \dots / Accepted \dots}

\abstract{We set out to determine the ratio, \qir, of rest-frame
8--1000-$\mu$m flux, \sir, to monochromatic radio flux, \s1.4, for
galaxies selected at far-infrared (-IR) and radio wavelengths, to
search for signs that the ratio evolves with redshift, luminosity or
dust temperature, \td, and to identify any far-IR-bright outliers --
useful laboratories for exploring why the far-IR/radio correlation
(FIRRC) is generally so tight when the prevailing theory suggests
variations are almost inevitable. We use flux-limited 250-$\mu$m and
1.4-GHz samples, obtained using \herschel\ and the Very Large Array
(VLA) in GOODS-North (-N).  We determine bolometric IR output
using ten bands spanning $\lambda_{\rm obs}= 24-1250\,\mu$m,
exploiting data from PACS and SPIRE (PEP; HerMES), as well as {\it
Spitzer}, SCUBA, AzTEC and MAMBO. We also explore the properties of an
\lir-matched sample, designed to reveal evolution of \qir\ with
redshift, spanning log \lir\ = 11--12\,L$_{\odot}$ and $z=0-2$, by
stacking into the radio and far-IR images. For 1.4-GHz-selected
galaxies in GOODS-N, we see tentative evidence of a break in the flux
ratio, \qir, at \l1.4\ $\sim 10^{22.7}$\,W\,Hz$^{-1}$, where active
galactic nuclei (AGN) are starting to dominate the radio power
density, and of weaker correlations with redshift and \td. From our
250-$\mu$m-selected sample we identify a small number of far-IR-bright
outliers, and see trends of \qir\ with \l1.4, \lir, \td\ and redshift,
noting that some of these are inter-related.  For our \lir-matched
sample, there is no evidence that \qir\ changes significantly as we
move back into the epoch of galaxy formation: we find \qir\ $\propto
(1+z)^{\gamma}$, where $\gamma=-0.04\pm 0.03$ at $z=0-2$; however,
discounting the least reliable data at $z<0.5$ we find $\gamma =
-0.26\pm 0.07$, modest evolution which may be related to the radio
background seen by ARCADE\,2, perhaps driven by $<$10-$\mu$Jy radio
activity amongst ordinary star-forming galaxies at $z>1$.}


\maketitle

\section{Introduction}

For samples of local galaxies -- on galactic and $\sim$100-pc scales
-- there is a good correlation between far-IR and radio emission
\citep{dejong85, helou85, condon91, yun01}. The correlation spans many
orders of magnitude in luminosity, gas surface density and photon,
cosmic-ray and magnetic energy density, and arises because the far-IR
and radio wavelength regimes share a common link with luminous,
massive stars and their end products -- dust, supernovae (SNe) and
cosmic rays. In the simplest models \citep[dubbed `calorimetry' --
e.g.][]{volk89, lvx96}, dust absorbs all of the ultraviolet radiation
from massive stars, re-radiating this energy in the far-IR, and when
those massive stars explode as SNe they generate cosmic-ray electrons
which lose all their energy in the radio regime, mainly via
synchrotron emission. A balance is thereby achieved between far-IR and
radio emission, assuming that the starburst timescale is sufficiently
long ($>$10$^7$\,yr).

Traditionally, \lir\ and \lrad\ are both employed to determine
star-formation rates, and the far-IR/radio flux density ratio has been
useful when estimating the redshift or \td\ of a distant starburst, or
when defining samples of AGN \citep{condon92, cy99, ivison02, bell03,
  chapman05, donley05}, or probing magnetic field strength
\citep{thompson06}.  For these reasons, and because of recent
observational advances at both far-IR and radio wavelengths, there has
been a deluge of FIRRC-related work recently, exploring why the
correlation exists and whether it continues to hold at progressively
larger look-back times \citep{garrett02, appleton04, ibar08,
  seymour09, ivison10, sargent10}. Prevailing theory
\citep[e.g.][]{lacki10} suggests that variations in the far-IR/radio
flux ratio should be virtually unavoidable and that the FIRRC thus
arises due to a mysterious combination of effects involving
bremsstrahlung, inverse Compton cooling, ionisation and the relative
fractions of primary/secondary cosmic-ray electrons/protons, as well
as the critical synchrotron frequency.

Aside from the modelling work of \citeauthor{lacki10}, recent advances
in this field have included the use of luminosity-matched samples
(between high and low redshift) to better probe evolution with
look-back time \citep{sargent10b} and the use of measurements spanning
the far-IR and radio wavebands to avoid assumptions relating to $k$
corrections \citep{ivison10}, although \citet{calzetti10} have argued
that bands beyond 24\,$\mu$m contain a contribution from dust heated
by stars from previous episodes of star formation and so we might not
necessarily expect the correlation to improve. In this paper we
introduce flux-limited 250-$\mu$m- and 1.4-GHz-selected samples of
galaxies from \herschel\ and the VLA, as well as a luminosity-matched
sample selected at 24\,$\mu$m, spanning $z=0-2$, and determine their
spectral energy distributions (SEDs) spanning the entire far-IR
spectral region.  We then investigate the FIRRC from the perspectives
of the 24-, 250-$\mu$m- and radio-selected samples.

\section{Sample selection and data analysis}

In this paper we present results from observations with \herschel\
\citep{pilbratt10}.  The SPIRE instrument, its in-orbit performance,
and its scientific capabilities are described by \citet{griffin10},
and the SPIRE astronomical calibration methods and accuracy are
outlined in \citet{swinyard10}. PACS is described by
\citeauthor{pog10} et al.\ (2010).

Our datasets are drawn from the common area observed by PACS and SPIRE
at 100, 160, 250, 350 and 500\,$\mu$m as part of HerMES\footnote{\tt
hermes.sussex.ac.uk} (Oliver et al., in preparation) and PEP (Lutz et
al., in preparation) in the GOODS-N field, prior to acquisition of
data for GOODS-{\it Herschel}. GOODS-N has also been observed with the
VLA at 1.4\,GHz \citep[1.7$''$ {\sc fwhm} --][]{biggs06, morrison10}
and {\it Spitzer} at 24, 70 and 160\,$\mu$m; we make use of these
data, as well as the 850-, 1100- and 1250-$\mu$m images of
\citet{borys03}, \citet{perera08} and \citet{greve08}.

\begin{figure}
\centerline{\psfig{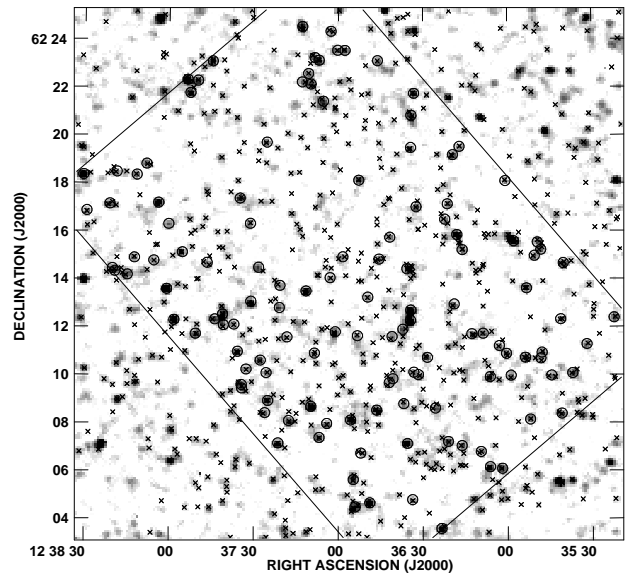}}
\caption{250-$\mu$m image of GOODS-N, with
    $\ge$5-$\sigma$ 250-$\mu$m [radio] sources marked by
    circles [crosses]. The PACS region 
    is also indicated. Here, $\sigma$ includes instrumental and
    confusion noise combined in quadrature, where $\sigma_{\rm
    conf}/\sigma_{\rm instr}\sim 5$ (cf.\ $\sim$2 for
    BLAST).}
\label{250map}
\end{figure}

We employ three GOODS-N galaxy samples, all selected above a signal-to-noise
threshold of 5\,$\sigma$:

\begin{enumerate}
\item 
128 galaxies selected at 250\,$\mu$m, without priors,
with $S_{\rm 250\mu m} \gs 20$\,mJy (Fig.~\ref{250map};
Smith et al., in preparation);
\item
247 galaxies selected at 1.4\,GHz (Fig.~\ref{250map})
with a \s1.4\ limit of $\sim$20\,$\mu$Jy, 137 with spectroscopic redshifts
\citep{barger08}, the remainder with photometric redshifts ($\left\langle
z \right\rangle = 0.94$; interquartile $z$, 0.56--1.76);
\item
a \lir-matched sample of 652 sources spanning $z=0-2$, selected initially
at 24\,$\mu$m \citep{magnelli09, berta10} then filtered to cover only
the decade of \lir\ between 10$^{11}\rightarrow 10^{12}$\,L$_{\odot}$
(LIRGs), where \lir\ is determined using
the models of \citet{ce01}.
\end{enumerate}

Far-IR and submm flux densities for the three samples are determined
using images convolved with appropriate point spread functions.  \sir\
is calculated by integrating under the well-sampled SEDs. Monte-Carlo
simulations are used to assess the uncertainty in \sir. The formal
error on $S_{\rm 24\mu m}$ was boosted by 3$\times$ to account for the
uncertain shape of the SED between rest-frame 8--70\,$\mu$m. A
modified blackbody fit to the measurements beyond 24\,$\mu$m (with the
emissivity index, $\beta= 1.5$) was used to determine \td.

For sample (1), additional procedures are implemented to define a
clean sample, free from blends: following the procedure of
\citet{downes86}, 107/128 sources are found to have secure ($P<0.05$)
radio identifications (ids) within a search radius, $r=10''$; we
discard the remainder. To avoid using those sources most severely
affected by blending, we further discard those with more than one
radio emitter within $r$, leaving 65 sources.  Of the galaxies without
a secure radio id, three have no plausible radio ids within $r$: a
potentially interesting sub-sample.  Measurements are made at the
radio positions for the 65 sources with secure, unambiguous ids, and
at the 250-$\mu$m positions for the three sources without radio
emission.

For sample (2), far-IR and submm measurements are made at the radio
positions.

For the \lir-matched galaxies (sample 3), median stacking is used to
measure \sir\ and \s1.4: we follow the procedure outlined by
\citet{ivison10}. Fluxes are calculated from 31$^2$-pixel$^2$ stacked
images in the ten available filters and \sir\ is determined as before.

\section{Results and conclusions}

\begin{figure}
\centerline{\psfig{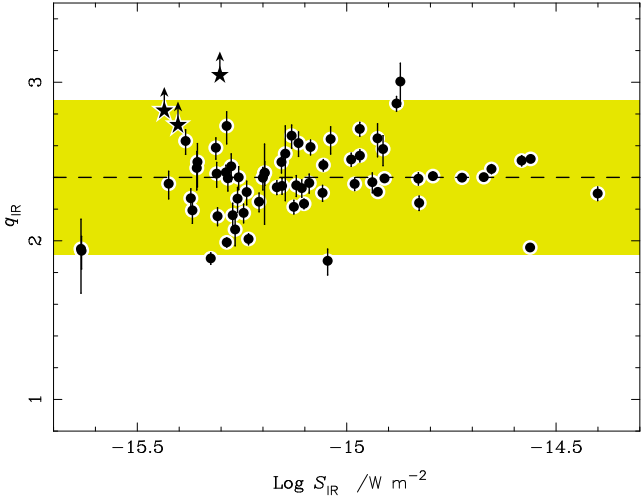}}
\caption{\qir\ versus \sir\ for those 250-$\mu$m-selected galaxies
(sample~1) with secure, unambiguous radio ids.  Those without
plausible radio ids are plotted as stars. The dashed line is the
median, \qir\ = 2.40; the shaded region represents $\pm$2$\sigma_q$
($\sigma_q=0.24$).}
\label{q250}
\end{figure}

\begin{figure}
\psfig{figure=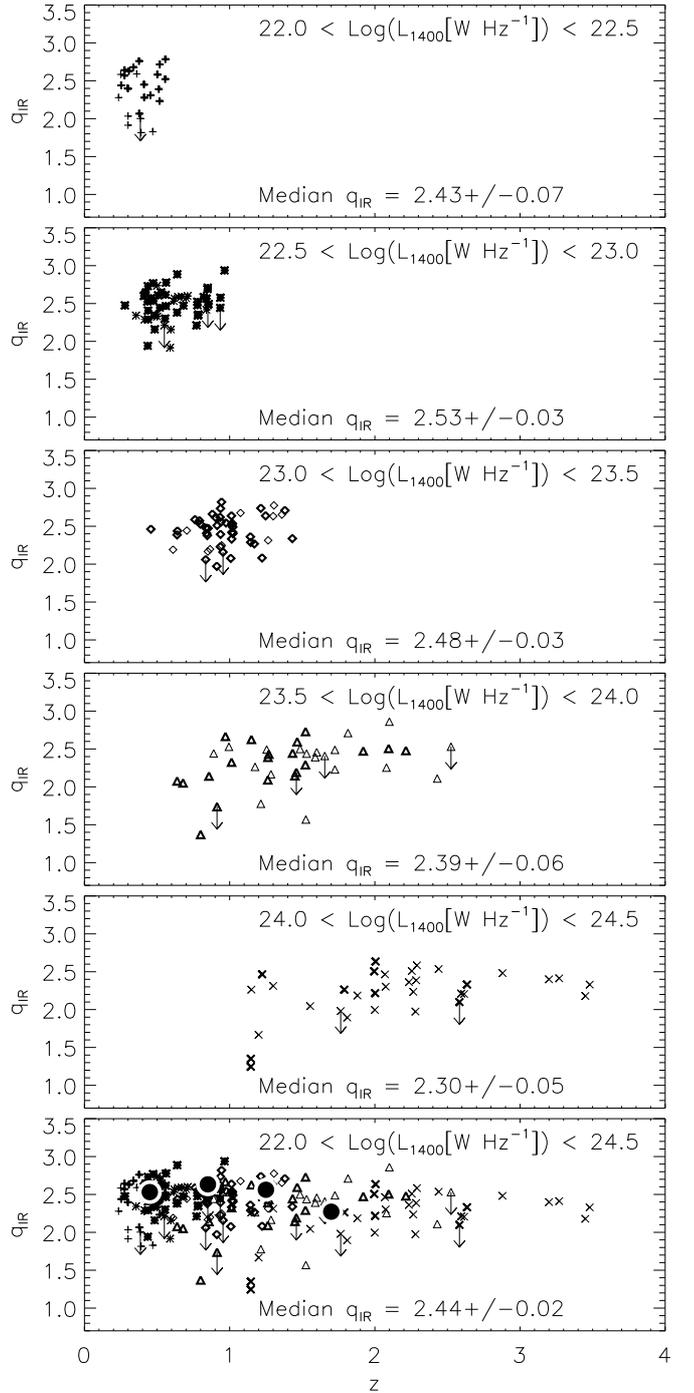,width=3.45in,angle=0}
\caption{\qir\ versus redshift for our radio-selected galaxies (sample~2),
in five bins of $K$-corrected \l1.4, plus the full sample.
Values of \qir\ for sample 3 are shown (as circles) for comparison.}
\label{qlrad}
\end{figure}

\begin{figure}
\psfig{figure=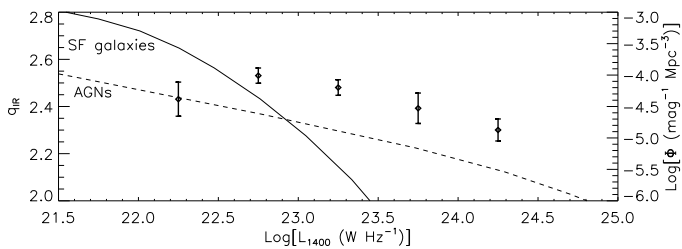,width=3.55in,angle=0}
\caption{Median \qir\ versus \l1.4.
The local luminosity functions of starbursts and AGN are shown
\citep{ms07}.}
\label{ql1400}
\end{figure}

\qir\ as utilised here is the logarithmic ratio of the rest-frame
8--1000-$\mu$m flux, \sir, and the 1.4-GHz flux density, $S_{\rm
1.4GHz}$, such that \qir\ = log$_{10}$ [(\sir/$3.75\times
10^{12}$\,W\,m$^{-2}$)/($S_{\rm 1.4GHz}$/W\,m$^{-2}$\,Hz$^{-1}$)], where
$S_{\rm 1.4GHz}$ is $k$-corrected assuming $S_{\nu}\propto
\nu^{\alpha}$, with $\alpha=-0.8$.

We begin with sample (1),  those selected at 250\,$\mu$m: \qir\ is
not a strong function of \sir\ (Fig.~\ref{q250}), nor of \s1.4. We see
no evidence of contamination by radio-loud AGN, consistent with the
findings of \citet{yun01}. Some galaxies stand out as potentially
far-IR-bright: these include the three galaxies without plausible
radio ids, two of which are detected at 70 and/or 160\,$\mu$m, so are
likely at low redshift with their radio emission resolved away.

Only 39/65 sources with unambiguous radio ids have redshifts (20
photometric, 19 spectroscopic; $\left\langle z \right\rangle = 0.98$;
interquartile $z$ = 0.46--1.52, similar to sample~2).  Nevertheless,
this sub-sample allows us to explore correlations between \qir\ and
luminosity, redshift and \td. We find significant ($>$95\% confidence
-- Table~\ref{trends}) trends for lower \qir\ amongst the most radio-
and far-IR-luminous galaxies, and the warmest and most distant, though
these parameters are likely inter-related. The dependence of \qir\ on
\l1.4\ is the strongest and likely reflects the influence of
low-radio-power AGN, of which more later; that of \qir\ on \lir\ is
more puzzling, perhaps reflecting the dependence of \lir\ on redshift
and/or \td\ \citep[e.g.][]{chapman05}, or selection effects (since
this trend is not seen for sample 2 -- see Table~\ref{trends}).

\begin{table}
\caption{Trends.}
\label{trends}
\centering
\begin{tabular}{l c c}
\hline\hline
\qir\ trend& Spear-    &Signifi-\\
           & man $\rho$&-cance\\
\hline
\multicolumn{3}{l}{Sample 1 (250-$\mu$m-selected galaxies with redshifts):}\\
(5.02$\pm$0.18) -- (0.105$\pm$0.008)\,log\,\l1.4 &$-0.48$&99.8\%\\
(6.09$\pm$0.33) -- (0.092$\pm$0.008)\,log\,\lir  &$-0.32$&95.6\%\\
(2.61$\pm$0.02) -- (0.081$\pm$0.007)\,(1+$z$)    &$-0.33$&96.0\%\\
(2.76$\pm$0.03) -- (0.008$\pm$0.001)\,\td\       &$-0.33$&96.1\%\\
\hline
\multicolumn{3}{l}{Sample 2 (radio-selected galaxies with redshifts):}\\
(4.92$\pm$0.21) -- (0.101$\pm$0.009)\,log\,\l1.4 &$-0.27$&99.9\%\\
(2.74$\pm$0.35) -- (0.007$\pm$0.009)\,log\,\lir  &$+0.07$&69.1\%\\
(2.55$\pm$0.02) -- (0.047$\pm$0.010)\,(1+$z$)    &$-0.15$&96.6\%\\
(2.60$\pm$0.02) -- (0.002$\pm$0.001)\,\td\       &$-0.16$&89.2\%\\
\hline
\end{tabular}
\end{table}

Fig.~\ref{qlrad} shows \qir\ versus redshift for our radio-selected
galaxies (sample 2), split into five log-spaced bins of \l1.4. Does
\qir\ evolve with redshift? One might conclude that it does, based on
the bottom panel of Fig.~\ref{qlrad}, where \qir\ $\propto
(1+z)^{\gamma}$, with $\gamma=-0.05\pm 0.01$
(Table~\ref{trends}). However, we must be aware of some strong
selection effects which make this evidence unreliable: radio emission
can be due to an AGN and several radio-loud objects with low values of
\qir\ are obvious in Fig.~\ref{qlrad}. Such AGN are more common at
$z\sim 2$ than today \citep[e.g.][]{wall05}; moreover, radio emission
from faint starbursts (with $\alpha=-0.8$, although see
\citealt{ibar10}) becomes more difficult to detect at higher
redshifts, such that the fraction of radio-loud AGN in a flux-limited
sample will rise, driving down \qir.  Indeed, Fig.~\ref{ql1400} shows
tentative evidence of a break in $\left\langle q_{\rm IR}
\right\rangle$ at \l1.4\ $\sim10^{22.7}$\,W\,Hz$^{-1}$. One might also
expect radio-loud objects (those with low \qir) to contain warmer,
AGN-heated dust, giving rise to the weak trend (89.2\% confidence --
Table~\ref{trends}) of decreasing \qir\ with increasing \td.

\begin{figure}
\centerline{\psfig{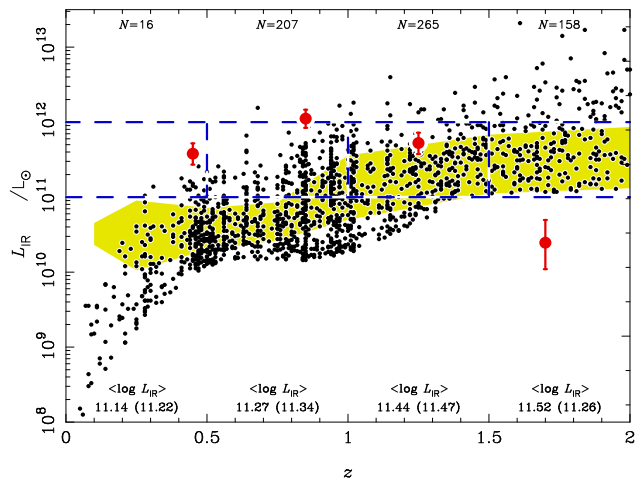}}
\caption{\lir\ (dots; left axis) and \qir\ (red circles; right axis)
-- the former determined via the models of \citet{ce01} -- versus
redshift for our \lir-matched sample.  The luminosity bounds and
redshift bins (dashed lines), the number of galaxies in each bin and
their predicted (measured) $\left\langle {\rm log} L_{\rm
IR}\right\rangle$ and measured $\left\langle q_{\rm IR}\right\rangle$
are all shown.  The shaded area represents a $\pm 1\sigma$ prediction
for \qir\ \citep{swinbank08, ivison10}.}
\label{lirz}
\end{figure}

Finally, we turn to our \lir-matched galaxies (sample 3), illustrated
in Fig.~\ref{lirz}.  The $\delta z=0.5$ bins provide significant
numbers of objects at near-constant \lir\ spanning $z=0-2$. As well as
being matched in \lir, there is another key difference between our new
sample and that used by \citet{ivison10}: although the new sample is
based initially on a flux-limited 24-$\mu$m catalogue, the final
selection is based on \lir, with model-dependent extrapolations from
the mid-IR (accurate to $\ls2\times$ across all bins --
Fig.~\ref{lirz}). This should lead to less contamination by AGN at the
blue end of the rest-frame 8--1000-$\mu$m band, where the relative
contribution to \sir\ can be substantial \citep[Figure~11
--][]{ivison10}. Using our new sample, there is no strong evidence
that \qir\ changes as we move back into the epoch of galaxy formation
at $z\sim 2$, with $\gamma=-0.04\pm 0.03$ where \qir\ $\propto
(1+z)^{\gamma}$, consistent with the findings of \citet{sargent10b}.
If we discount the $z<0.5$ data, which comprise only 16 galaxies which
are not well matched in \lir\ to the higher redshift bins, we find
$\gamma = -0.26\pm 0.07$. This is similar to the $\gamma = -0.15\pm
0.03$ found by \citet{ivison10} who noted reports that evolution in
\qir\ could be related to the radio background seen by ARCADE\,2
\citep{fixsen10, seiffert10}. Our sample, with $\left\langle S_{\rm
1.4GHz} \right\rangle \ls 10\,\mu$Jy at $z\gs 1$, is consistent with
the idea that evolution of the FIRRC might be driven by $<$10-$\mu$Jy
radio activity amongst ordinary star-forming galaxies at $z>1$
\citep{singal10}.

\begin{acknowledgements} The data presented in this paper will be
released through the {\it Herschel} Database in Marseille HeDaM ({\tt
  hedam.oamp.fr/HerMES}).  SPIRE has been developed by a consortium of
institutes led by Cardiff Univ.\ (UK) and including Univ.\ Lethbridge
(Canada); NAOC (China); CEA, LAM (France); IFSI, Univ.\ Padua (Italy);
IAC (Spain); Stockholm Observatory (Sweden); Imperial College London,
RAL, UCL-MSSL, UKATC, Univ.\ Sussex (UK); Caltech, JPL, NHSC,
Univ.\ Colorado (USA). This development has been supported by national
funding agencies: CSA (Canada); NAOC (China); CEA, CNES, CNRS
(France); ASI (Italy); MCINN (Spain); SNSB (Sweden); STFC (UK); and
NASA (USA).  PACS has been developed by a consortium of institutes led
by MPE (Germany) and including UVIE (Austria); KUL, CSL, IMEC
(Belgium); CEA, OAMP (France); MPIA (Germany); IFSI, OAP/AOT,
OAA/CAISMI, LENS, SISSA (Italy); IAC (Spain). This development has
been supported by the funding agencies BMVIT (Austria), ESA-PRODEX
(Belgium), CEA/CNES (France), DLR (Germany), ASI (Italy), and
CICYT/MCYT (Spain).
\end{acknowledgements}

\bibliographystyle{aa}
\bibliography{14552}

\end{document}